\documentclass[]{agujournal2019}
\usepackage{url} 

\journalname{Space Weather}
\begin{document}

\title{Addressing Known Challenges in Solar Flare Forecasting I:  Limb-Flare
Prediction with a $4\pi$ Full-Heliosphere Framework}

%
%

\authors{K.~D. Leka\affil{1}, Eric L. Wagner\affil{1}, Lisa Upton\affil{2}, Bibhuti Kumar Jha\affil{2},
Kiran Jain\affil{3}, Sara Petty\affil{1}}

\affiliation{1}{NorthWest Research Associates, 
Boulder, Colorado, USA.}
\affiliation{2}{Southwest Research Institute, Boulder, Colorado, USA.}
\affiliation{3}{National Solar Observatory, Boulder, Colorado, USA.}

\correspondingauthor{K. D. Leka}{leka@nwra.com}


\begin{keypoints}
\item Solar flare forecasting generally fails to forecast flares near or just-beyond the solar limb.
\item We present a new full-heliosphere forecasting method that combines a surface flux-transport model and far-side helioseismology.
\item Within this proof of concept framework, geo-effective limb flares were forecast with demonstrably improved performance.
\end{keypoints}

%
%

\begin{abstract}
A demonstrated failure mode for operational solar flare forecasting
is the inability to forecast flares that occur near,
or just beyond, the solar limb.  To address this shortcoming, we
develop a ``$4\pi$'' full-heliosphere event forecasting framework and
evaluate its statistical classification ability against 
this specific challenge.  A magnetic
surface flux transport model is used to generate full-sun maps of the
photospheric radial magnetic field from which active
regions (ARs) are identified and tracked using a new labeling scheme that 
is observer-location agnostic and allows for post-facto modifications.
Flare-relevant magnetic parameters couple to a ``visibility'' index that
specifies AR location relative to the visible solar limb and expected flare
detection.  Flare labels are assigned according to peak Soft X-ray
flux, and a statistical classification is performed using nonparametric
discriminant analysis. A version where new or emerging ARs on
the far (``invisible'') side of the Sun are incorporated into the model
by way of far-side helioseismology, is also tested.  We evaluate the new
framework by its performance specifically on the limb areas using Brier
Skill Score and ROC Skill Score, finding improvement at the $2\sigma$
level or less.  However, we do find that the number of False Negatives,
or ``missed'' forecasts decreases, and find strong evidence that the
additional information provided by the far-side helioseismology can help
predict near- and just-beyond-limb flares, particularly for East-limb events.  
While individual components
of this framework could be improved, we demonstrate that
a known failure mode for solar flare forecasting can be mitigated with
available resources.
\end{abstract}

\section*{Plain Language Summary}
Solar flares can occur anywhere there are concentrated magnetic fields
on the Sun (active regions or sunspot groups). Predicting solar flares
often uses the character of those magnetic fields to estimate the
likelihood of a flare occurring.  However, a significant fraction of solar
flares that are visible from the Earth (thus potentially hazardous
to our technological systems, especially communications) occur in regions
where the magnetic field data becomes increasingly noisy and unreliable,
or just beyond the edge of the Sun where we cannot see the magnetic
fields at all.  Forecasting for these particular flares is generally 
unsuccessful, thus identified as the ``Limb-Flare Challenge''.  We have
developed a framework that uses physics-based modeling to create a map of
the ``$4\pi$'' or ``full-Sun'' (including the far-side, or unobserved part of the Sun),
and test this new ability to address this specific forecasting challenge.
Additionally, we quantify the improvements made when additional
information is included from far-side helioseismology, a technique using
Earth-visible data to detect strong magnetic fields on the far side of
the Sun.  We show that with this new ``$4\pi$'' forecasting framework,
and with the additional information about the far-side magnetic field, the
number of limb-flares ``missed'' by forecasts can successfully decrease.


\section{Introduction}
\label{sec:intro}

The challenges of forecasting solar flares are well documented and
discussed \cite<e.g.>[]{EEG_chapter,ffc3_1,flarecast}.  Recently, it was
demonstrated that operational forecasting facilities can indeed provide positive
skill against random- or climatology-based forecasts, as measured with
multiple evaluation metrics \cite{ffc3_1}.  On the research side, there has
been a plethora of recent efforts to improve flare forecasts, the vast majority 
involve exploring machine-learning (ML) methods (34 refereed publications with titles
that include ``Solar Flare Forecasting'' in just 2024 and 2025 thus far), though
some investigate new data sources as well \cite{nci_aia,SunD_etal_2023,Zbinden_etal_2024}.
Central to this activity is the increasing sample size of data, in
large part from the Solar Dynamics Observatory facility \cite<SDO;>[]{sdo}
that has observed the visible solar disk almost continuously for
over 15 years at the time of this writing.  Of the SDO instruments,
the magnetic field data provided by the Helioseismic and Magnetic
Imager \cite<HMI;>[]{hmi,hmi_wavelength,hmi_cal,hmi_invert,hmi_pipe}
has been central to flare-forecasting research -- in part due to the 
``Space Weather Active Region Patch Parameters'' data product that is provided.  
This data product includes a small number of 
pre-computed parameters describing photospheric magnetic concentrations or 
HMI Active Region Patches
\cite<``HARPs''; >[]{hmi_sharps} that are readily available as meta-data
\cite{hmi_sharps}.  Interest in ML-based forecasting
methods has become widespread, with many efforts exploring the efficacy of 
combining the SHARP parameters with different flavors of ML tools; 
improvement in forecasting capability have been reported, although cross-publication
comparison of any particular metric must be done with care \cite<see discussions 
in>[]{allclear,Kubo2019,ffc3_1}.

One specific performance gap was recently identified as being
particularly difficult for present approaches to address: forecasting
limb flares \cite{ffc3_3}.  Due to the lack of photospheric magnetic
field data or even white-light imaging towards the edge of, or beyond, the
visible solar disk, algorithms generally cannot provide forecasts for
active regions (ARs) located (or even suspected to be located) in these areas.
Some facilities do produce extended validity-period forecasts, which could
provide forecasts for an AR -- but generally for no more than a day past the western 
limb-transit \cite<{\it e.g.}, the 2- or 3-day forecasts available from NOAA/SWPC, DAFFS,
and others; see>[for descriptions]{ffc3_1}.  However, this option
only addresses West-limb events.   Solar flares that occur at, or just
beyond, the visible East limb can also impact the Earth's ionosphere, 
for example, if the Soft X-Rays (SXR) and [Extreme] Ultraviolet (E/UV)
radiation from event-associated coronal loops are visible above the limb
\cite<see, {\it e.g.},>[]{Sept12flarealert}.

A recent effort has attempted to address this challenge with a forecasting
system based solely on images of E/UV emission, which can capture coronal loops 
that appear above the limb \cite{LeeJ_etal_2024}.  However, this approach still relies on 
Sun-Earth-line data, and requires active-region-related emission to be present
(meaning: an AR must be large enough, active enough, and close
enough to the limb that the associated coronal loops are available for analysis).  

Here we present a different approach: ``$4\pi$'' event forecasting.
The basic technique is to invoke a ``full-Sun'', meaning both 
Earth-facing ``visible'' and far-side ``not visible'' or a $4\pi$ representation of the solar
photospheric magnetic field.  We parameterize identified ARs
to quantitatively characterize their magnetic field distributions
(as is common for forecasting methods), and evaluate a sample of labeled
event-producing and event-quiet ARs using a statistical
classifier such that future regions and their parameters can be
labeled predictively.  In this new framework of $4\pi$ full-Sun forecasting, we
additionally (1) include information about far-side (invisible from the
Earth) ARs identified with far-side helioseismology, and (2)
provide forecasts for both ``occulted'' and near-limb regions as well as
Earth-facing regions for which forecasts would otherwise be unavailable.

We describe the full methodology in Section\,\ref{sec:methods},
including the framework and input (both observational and modeling,
near-side and far-side) to generate the $4\pi$-Sun in
Section\,\ref{sec:4piSun}, and 
describe a new AR referencing scheme for $4\pi$ analysis 
in Section\,\ref{sec:arnumber}.  We describe the parameterizations
in Section\,\ref{sec:parameters}, the 
statistical approach and specific tests designed and carried
out in Section\,\ref{sec:stats_general}, the results in 
Section\,\ref{sec:results} and finish with a discussion in
Section\,\ref{sec:disc}.

\section{Methodology}
\label{sec:methods}

The $4\pi$ forecasting framework consists of essentially three parts: generating the full
$4\pi$ solar surface magnetic field, identifying the ARs and parametrizing their 
magnetic field distribution, and a statistical analysis of those parametrizations in the context
of flare activity.

\subsection{Generating the $4\pi$ Sun}
\label{sec:4piSun}

The photospheric magnetic field on the full Sun ({\it vs.} the
visible disk) is required input for numerous heliospheric models,
including solar wind and heliospheric event propagation \cite<{\it e.g.},
the Wang-Sheeley-Arge-ENLIL solar wind and CME propagation model;>[]{Arge_etal_2004,Odstrcil_etal_2004}.  
This boundary input can be 
``built up'' by way of creating synoptic or synodic input over the course of 
the $\approx$1\,mo rotation period, but the information on 
the far side of the Sun is then necessarily very outdated.  For a synchronic $4\pi$ map that 
represents the Sun at a given moment, surface flux transport 
models have been developed to assimilate Earth-facing photospheric magnetic 
field information, and then evolve the field according to modeled physics.

\subsubsection{The Advective Flux Transport Model}
\label{sec:aft}

Here we use the Advective Flux Transport model
\cite<AFT; >[]{UptonHathaway2014b,UptonHathaway2014a} to construct synchronic
estimates of the solar photospheric magnetic flux over the entire Sun.  In brief, data
from Earth-facing sources are assimilated; in the present
version HMI line-of-sight magnetic maps from the {\tt hmi.M\_720s}
data series are used to estimate the radial component $B_r$ using a
``$\mu$-correction'' \cite{Svalgaard_etal_1978,WangSheeley1992};
\cite<see also>[for discussion]{bbpot}.  The assimilation window is a
weighted function that falls off radially from the center of the disk. 
This has the effect of de-emphasizing data near any visible solar limb, where 
magnetic field measurements are known to be problematic.

AFT then solves the radial component of the induction equation:
\begin{equation}
\frac{\partial B_r}{\partial t} + \nabla \cdot (\vec{u} B_r) = S(\theta,\phi,t) + \eta \nabla^2 B_r
\label{eqn:induction}
\end{equation}
\noindent
where $B_r$ is the radial magnetic field, $\vec{u}$ denotes horizontal
flows including convective and axisymmetric flows (differential rotation and
meridional circulation), $S$ is a magnetic source term (representing new flux added to the model), and $\eta$ is 
a numerical diffusivity term. In our case, the induction equation is solved in 15\,min time steps with HMI data 
assimilated hourly. The convective velocities are implemented using a spherical harmonic code that simulates 
cellular convective structures (of supergranular spatial scales and larger) that evolve in time. These convective 
cells have lifetimes that are inversely proportional to their sizes and they are advected with observed 
axisymmetric flows \cite{UptonHathaway2014b} . This convective simulation is used to generate a series of 
latitudinal and longitudinal flow maps with a 15-minute cadence, coinciding with the model time step. AFT maps 
generated with this configuration are referred to as AFT ``Baseline'' maps.

The $4\pi$ AFT maps are sampled at $0.352^\circ$ in both latitude and longitude
(corresponding to a map size of $512 \times 1024$ pixels), which is finer than most
surface flux transport models,
especially when serving as the inner boundary-input to global models.  Here, we are
in fact interested in the small-scale behavior and characteristics of 
ARs, hence the higher AFT spatial sampling is key.  The AFT magnetic field maps are
field-filled (unity fill fraction), and the magnitude of the signal
is referred to in units of Gauss-equivalent Mx\,cm$^{-2}$ to reflect this characteristic (and to 
be consistent with HMI data.  For this experiment, we use AFT-generated maps created at a cadence of
three times each day (00:00, 08:00, and 16:00 UT).

\subsubsection{AFT Active-Region Detection}
\label{sec:aft_ardetect}

Active regions are the dominant source of all flares, particularly for
larger flares.  As such, we apply the following algorithm to detect
concentrations of magnetic field in the AFT maps, which we denote as AFT active
regions (AFT-ARs). This approach is an extension of the method proposed by
 \citeA{sreedevi_autotab_2023}.

First, an intensity threshold of $150$\,Mx\,cm$^{-2}$ is applied to the image and
pixels below this threshold are excluded.  Second, a morphological
closing operation, with a circular kernel of 4 pixels, is applied to fill the gaps 
between pixels separated by this threshold; this step also helps to join the positive and negative polarities
of an AR bipole.  Next, an area threshold of 20\,pixels
(roughly $4.5^\circ\,^2$ or approximately $6.6{\rm\, Mm}^2$) is applied to
exclude very small regions, based on the lower limits found using a 
frequency-distribution analysis of the 
regions detected to this point.  Arguably this step may remove the very initial detection 
of growing regions, but the dominant effect is to remove small spurious detections.  
Finally, a moderate flux balance condition and flux
threshold are applied primarily to eliminate plage regions and similar unipolar
areas. Here, flux balance
is defined as:
\begin{equation}
r = \frac{|(\Phi^+ + \Phi^-)|}{(\Phi^+ - \Phi^-)}
\label{eqn:fluxbalance}
\end{equation}
\noindent
with $\Phi^+, \Phi^-$ being the total signed magnetic flux of positive and negative
polarities, respectively.  As such, a flux-balanced AR has $r=0$ and 
fully unbalanced region (unipolar) produces $r=1$.  For the present rendition, we require
an AR to have $r < 0.7$ (requiring no worse than $5\times$ imbalance in 
flux between $\Phi^+$ and $\Phi^-$), and $\Phi > 5\times10^{19}$\,Mx.

\begin{figure}
\centerline{
\includegraphics[width=1.0\textwidth,clip, trim = 10mm 2mm 4mm 8mm, angle=0]{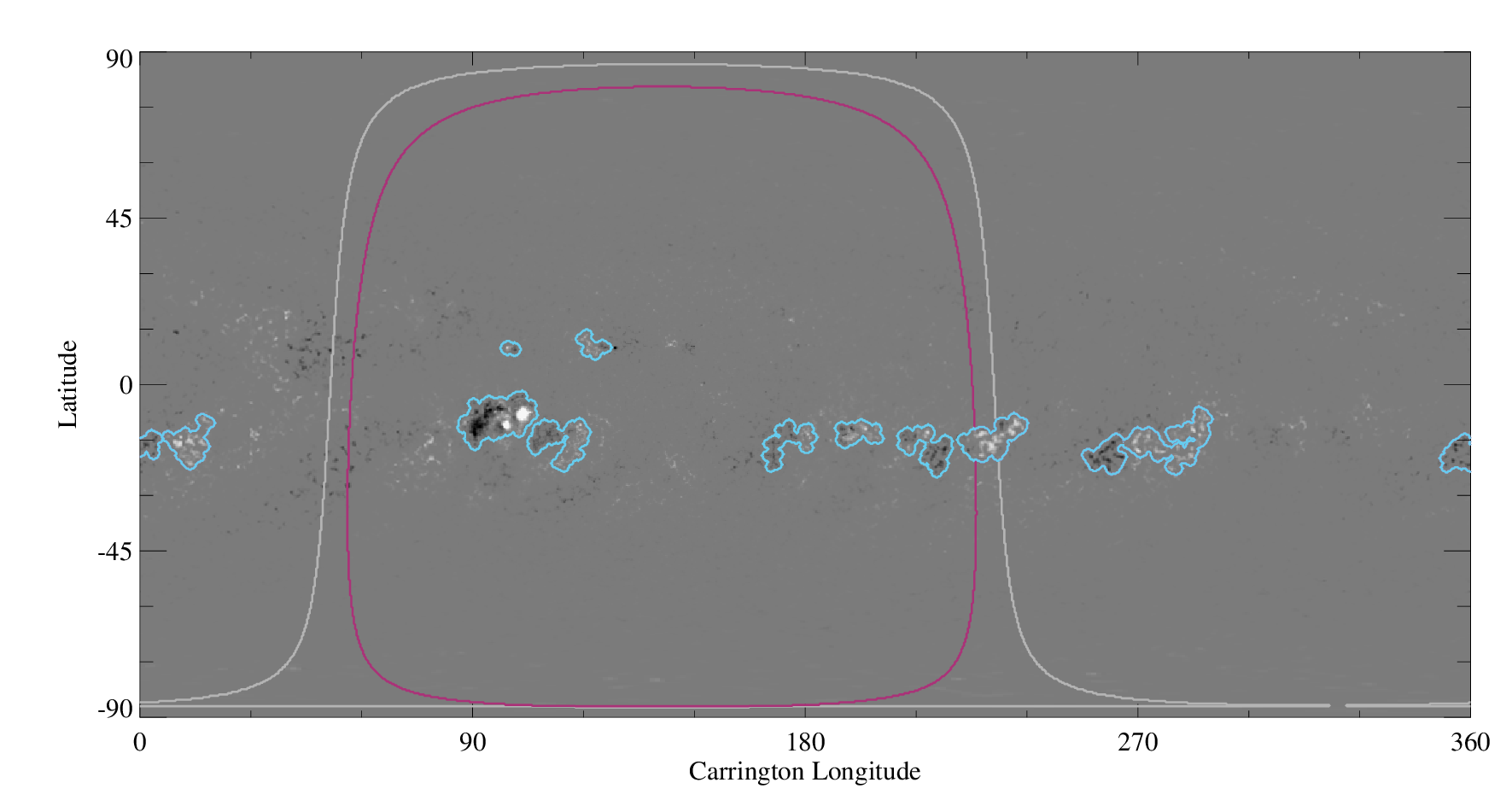}}
\caption{An example AFT full-Sun (or ``4$\pi$'') magnetic field map on 2014.01.04 18:00\,UT, 
scaled to $\pm 1,000\,{\rm Mx\,cm}^{-2}$, with the solar limb (grey), the AFT assimilation window 
(magenta) and identified AFT active regions, or AFT-ARs (blue).  Note that the 
b0-angle on this day
is such that the south pole is slightly visible.  In this example, the regions fully within the assimilation
window would be assigned {\tt Disk}, the two regions on the far-side (at CL=[$0^\circ,280^\circ$])
would be assigned {\tt Far}, and the region centered at ${\rm CL}\approx230^\circ$ that spans 
both the assimilation and limb terminators would be assigned {\tt Occ}, or ``occulted'' (see Section~\ref{sec:vis} for 
details). } 
\label{fig:aft_eg}
\end{figure}

An example AFT map with identified regions is shown in Figure~\ref{fig:aft_eg}.

\subsubsection{Far-Side Magnetograms from Far-Side Helioseismology}
\label{sec:gong}

We have no direct and consistent observations of the Sun's magnetic field beyond
the limb  of the Sun as seen from the ``near-side'', or the Sun-Earth line.  
However, the Sun's magnetic field does evolve and
new ARs continue to emerge on the ``invisible'' far-side of the Sun.  Helioseismology can 
be used to infer activity that involves significant magnetic flux
on the far side \cite{cl_dcb2000,dcb_cl2001,igh2007,igh2010}. For this experiment, we use helioseismic 
output (phase-shift maps) from the far-side processing pipeline
of the Global Oscillation Network Group \cite<GONG;>[]{Harvey1996, jain2021}. 
Each far-side phase-shift map is constructed by combining
five successive phase-shift maps taken at 6\,hr intervals over a 48\,hr period.
Twice-daily phase-shift maps at 00:00UT and 12:00UT are 
then converted to 
magnetic flux-density maps following \citeA{igh2007,igh2014,MacDonald_etal_2015} on 
a grid of $180\times360$ pixels.  The information on these maps are used to inform the 
AFT model of additional far-side magnetic flux emergence (as described below).

\subsubsection{Far-Side Active Region Detection}
\label{sec:farside_ars}

To detect magnetic activity in far-side magnetic flux maps derived from GONG phase-shift data,
we use a multi-step approach:
\begin{enumerate}
\item We first define a flux-density threshold for the selected GONG inferred flux-density 
maps using the relation $I_{\rm th} = \overline{I} + 3\sigma(I)$, where $\overline{I}$ and $\sigma(I)$
are calculated only over pixels where the inferred flux density is more
than 0.1 Mx\,cm$^{-2}$ (this is to avoid inversion-generated noise values).  This is 
performed using the sigma clipping method {\it i.e.}, iteratively calculating 
the mean by removing the intensity values beyond $3\sigma$ \cite{sigmaclip}. 
Typical threshold values are 
of order 90\,Mx\,cm$^{-2}$, with $\sigma(I) = 20$ after the sigma clipping.
Any feature in the flux-density maps that exceeds this flux-density threshold
is labeled as a candidate region.

\item We next ensure the persistence of a candidate region by examining
the signal in the GONG magnetic flux map in either the previous observation
or the next.  To do this, candidate regions are first identified in these 
neighboring maps using a slightly relaxed threshold, {\it i.e.} $I_{\rm th} = \overline{I} + 2\sigma(I)$. 
We then check for region overlap, and if the regions from either of the 
adjacent GONG flux-density maps overlap by more than 5 pixels, we confirm the persistence of 
the candidate region. 

\begin{figure}
\centerline{
\includegraphics[width=0.75\textwidth]{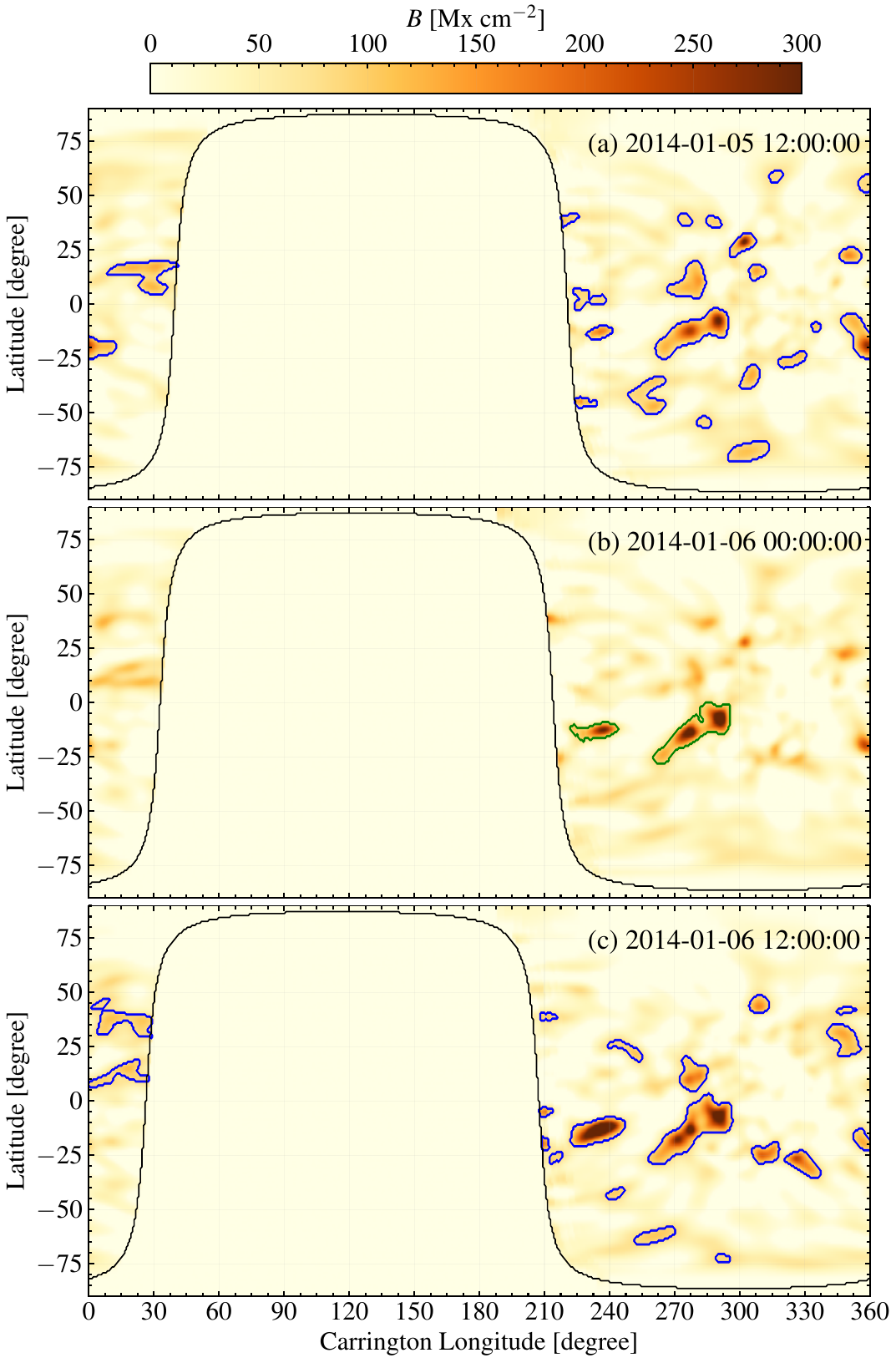}}
\caption{
GONG flux-density maps for a target time (panel b, middle) and the prior map 12 hr prior (panel a, top) and
the subsequent map 12hr later, scaled as shown.
Blue contours (panel (a) at -12h and panel (c) at +12hr) identify the potential AR regions
based on the relaxed threshold (see text for details); green contours in panel (b) 
identify the final set of far-side ARs 
that satisfy all the criteria for far-side ARs including the more stringent threshold applied for 
the target time to avoid noise.}
\label{fig:gong_AR}
\end{figure}

\item Once we identify a persistent candidate AR, we calculate its properties including 
latitude and Carrington longitude based on the flux-density-
weighted spatial mean (a magnetic center-of-mass or MCOM approach), total magnetic flux, 
and maximum flux-density.  
We only accept candidate far-side ARs with a peak magnetic flux density of at least 250\,Mx\,cm$^{-2}$, 
total magnetic flux greater than $5\times10^{20}$\,Mx and centroid latitude within the 
limits of the active zone defined as:
\begin{equation}
\ell= \ell_0 \exp\left(-\frac{T-T_0}{\tau}\right)\pm20^\circ, ~\ell_0=\pm 28^\circ~\&~\tau=90;
\label{eqn:active}
\end{equation}
where $\ell_0$ and thus $\ell$ is a signed latitude (that is, referring to both hemispheres), and 
$T-T_0$ is the number of months from the beginning of the solar cycle 
\cite{Hathaway2011a}.  The relevant parameters have been tuned to make the 
distribution of identified far-side ARs (in latitude and number) as consistent as possible to those
 observed on the near side of the Sun.  The choice of $20^\circ$ is based on the 
maximum width of the ``butterfly wings'' observed in near-side AR data over a complete solar 
cycle.  To ensure robustness, a limit corresponding to $4$ times the $1\sigma$ value 
in \citeA{Hathaway2011a} Eqn.~9 was adopted.
\end{enumerate}

Figure~\ref{fig:gong_AR} (panels a, b, and c) provides
a representative example of the process.  The three panels
show three consecutive far-side flux-density maps: Figure~\ref{fig:gong_AR}b represents
the target map at the time of interest for identifying far-side ARs and Figures~\ref{fig:gong_AR}a, c
are the far-side flux-density maps at $\pm12$hr, used to verify the consistency of the ARs.
Contours at all three times represent region candidates although candidates
at the neighboring times ($\pm12$hr) are based on the relaxed threshold.  The only AR candidates that
satisfied all of the criteria are those with contours shown in Figure~\ref{fig:gong_AR}b, at the target time.

\subsubsection{Far-Side Active Region Information Incorporation into AFT}
\label{sec:gong2aft}

Once the far-side ARs are identified on the GONG flux-density maps, the information 
needs to be incorporated in the AFT model.  The steps for this are as follows:

\begin{enumerate}
\item  A ``region mask'' is defined for each candidate AR by the contours described above.  
Because far-side AR candidates display considerable fluctuation in their location, 
we extend the mask by 10 pixels (10$^\circ$) in longitude both East/West and 5 pixels (5$^\circ$) 
in latitude both north/south.  This 
mask is used to calculate the total flux in the same region using the already-modeled AFT map.
If the total magnetic flux in the mask (for the far-side AR candidate) 
from the GONG flux-density map is greater than the total magnetic flux of the 
AFT map in same mask, we start the process of adding the far-side AR into AFT; otherwise it is excluded. 
This step is performed to ensure that we only add in new far-side ARs or flux from ARs 
with significant growth.  Any subsequent AR decay is completely governed by the surface 
flux transport processes.

\item To add a candidate GONG-identified AR to AFT, 
we use the AR's location and magnetic flux as derived from the 
GONG flux-density maps to construct an ideal bipolar region modeled
as two 2-D Gaussian distributions of opposite-polarity magnetic flux, with 
standard deviation and amplitude tuned to match characteristics of HMI magnetograms.  
These simple bipoles are provided with 
the expected Joy’s law tilt according to latitude and Hale’s law for polarity orientation 
according to hemisphere \cite{Hale_etal_1919}. The flux {\it vs.}
area relation \cite{Sheeley66, Mosher1977}, and area {\it vs.} polarity separation distance 
relation \cite{HathawayUpton2016}, are used to determine the size and 
separation of the negative/positive polarity flux concentrations. 

\item Tests indicated that simply using the estimated magnetic flux density 
from the GONG maps directly added too much flux to the AFT maps,
as evidenced by the behavior of newly-inserted far-side ARs undergoing a sharp decrease in
magnetic flux density once the AFT data-assimilation process corrected them with HMI data.
As such, we calculate the fraction $f$ of total magnetic flux from a 
GONG-identified AR that needs to be added into AFT with a new bipole, as
\begin{equation}
f = \frac{\alpha\Phi_{\rm Far} - \Phi_{\rm AFT}}{\alpha\Phi_{\rm Far}}.
\label{eqn:alpha}
\end{equation}
We have found upon initial analysis that the assimilation of newly inserted ARs
was smoothest when $0.25 \leq \alpha \leq 0.35$, and thus assigned $\alpha=0.3$.
However, the present work is focused on shorter-term AR evolution and in fact
$\alpha$ may be non-linear with magnetic flux density, AFT-based evolution prior to near-side 
assimilation, AR morphology, or even solar cycle phase.
Further investigation is on-going but beyond the scope of this application.
\end{enumerate}
The effect of adding far-side ARs, or adding magnetic flux to established ARs
as they transit the far-side of the Sun, is to (unsurprisingly) increase the number of AFT-ARs above what 
is identified by AFT-Baseline, particularly on the Eastern limb.

\subsection{Active-Region Tracking and Numbering}
\label{sec:arnumber}

NOAA assigns sequential numbers to their identified ARs, 
as do the ``HMI Active Region Patches'' \cite<HARPs>{hmi_sharps}, even though
the two facilities have very different definitions of what constitutes
an AR.  Common to these two numbering schemes is that a region 
which disappears beyond the West limb as the Sun rotates and transits the ``far side''  will be 
assigned a new number upon re-appearance on the East visible limb.  
This situation creates two sources of confusion: first, the same
concentration of magnetic flux is assigned to multiple identities over
its lifetime, and second, sequential numbering prohibits any 
incorporation of newly identified AR into the record, or 
even corrections made to the list {\it post facto}.
As this project necessarily tracks ARs possibly over numerous 
solar rotations and includes ARs that may be first identified 
on the far side, a new region-labeling system is needed.  

In this new ``$4\pi$'' framework, once magnetic regions are identified by the methods 
described above (Sections\,\ref{sec:aft_ardetect},~\ref{sec:farside_ars}),
they are assigned an identification and tracked using a modified AutoTAB (\texttt{autotabpy}) 
algorithm \cite{Jha2021, sreedevi_autotab_2023, autotab2025}. 
Newly identified ARs are labeled with a code (an ``AFT-ID'')
that specifies
the date of first detection and the MCOM latitude and Carrington longitude.  
As an example, the AFT-AR with the ID ``{\tt 20121109TN11243}'' (or ``AFT-ID'')
first became an
AFT-identified AR on 2012 November 09 at Latitude N11 Carrington
Longitude 243.  This identifier is an easily-parsable string and most
importantly, unique; it does not need to change according to the relative
location of the Earth.  After running on more than 10 years of data, this system
does not produce duplicates or conflicts and allows for modifications according to 
detection algorithm particulars, without jeopardizing the numbering scheme.

AFT-ARs are linked, then, to their corresponding NOAA~AR
numbers by way of the latter's Latitude and Carrington Longitude, within
a $10^\circ$ window.  As such, a single AFT-ID can have multiple
NOAA~ARs associated with it, not unlike the HARP-IDs.  AFT-ARs that are in close 
proximity are combined, or absorbed, into the biggest and/or 
oldest match -- again, within $10^\circ$.  This process effectively treats multiple AFT-ARs 
(as initially identified) that would match to an identical NOAA AR number as a single AFT-AR.
Additionally, since
AFT-ARs may in fact disappear and re-appear with a new AFT-ID (due to varying
noise levels and subtle challenges to temporal continuity), a single NOAA~AR may 
in fact be associated with more than one AFT-ID over the course of their
respective lifetimes (this is the case for NOAA-AR 12205, discussed in Section~\ref{sec:cases}, below).  
It is not evident in the available NOAA documentation
exactly when a numbered NOAA~AR that transits the full far 
side is assigned a new number 
once it appears at the East solar limb, but by associating each
unique AFT-AR with any and all NOAA~AR numbers, limb- and beyond-limb
events can be correctly labeled (see Section\,\ref{sec:eventlist}).

\subsection{Parametrization}
\label{sec:parameters}

The magnetic field distributions of the AFT-ARs are parametrized with a focus
on describing the character of the AR in the context of flare productivity.  
From experience developing {\tt DAFFS-G} \cite<>[see \url{www.nwra.com/DAFFS}]{nci_daffs,ffc3_1}
where flare prediction is based on the line-of-sight magnetic field
from GONG, the parameters used here include those that have shown efficacy
in this context.  It must be emphasized 
that we are {\it not} trying to achieve the ``best'' forecasts, but instead provide 
a ``proof of concept'' for 4$\pi$ forecasts.

\subsubsection{Identification, location, and visibility}\label{sec:vis}  The identification parameters
include the AFT-AR ID (transformed into a 32-bit integer code for the 
Fortran code in the statistical analysis, see Section~\ref{sec:nci}), 
and a code for whether the data were assimilated into AFT from the HMI input.
The location parameters include not just the MCOM Latitude / Carrington Longitude
of the AFT-AR, but the Stonyhurst Latitude / Longitude, extended to $\pm180^\circ$ 
($0^\circ$ being central meridian), and the cosine of the 
observing angle $\mu =\cos(\theta)=\cos({\rm Stony lat.}) \cos({\rm Stony Long.})$ such
that $\mu < 0$ denotes beyond-visible MCOM AFT-AR locations.  

\begin{figure}
\centerline{
\includegraphics[width=0.95\textwidth,clip, trim = 30mm 50mm 30mm 20mm, angle=0]{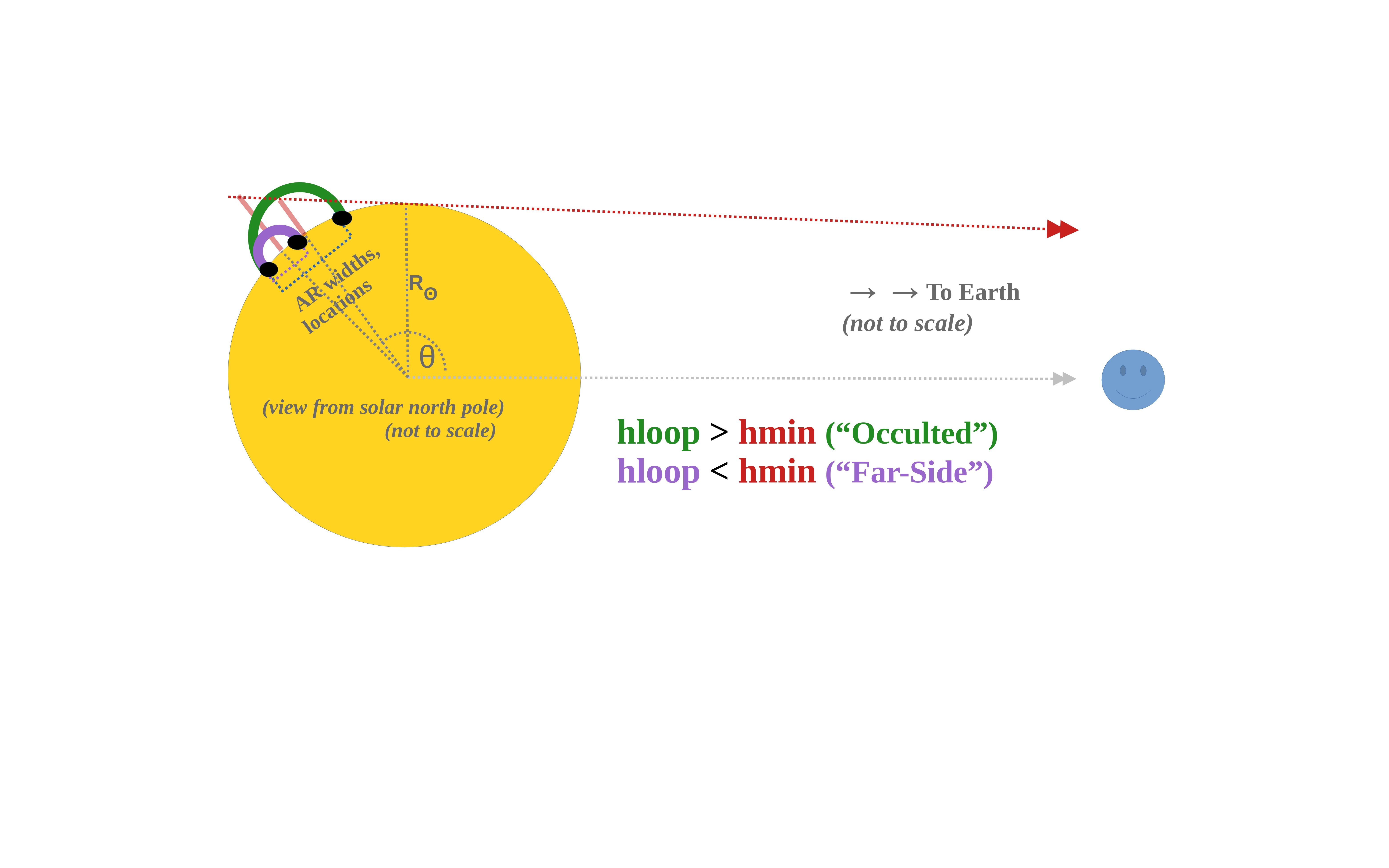}
}
\caption{
Cartoon representation (with absolutely no scale intended) of 
a far-side AR region (purple) and an occulted region (green), 
and the effective Stonyhurst longitude $\theta>90^\circ$ indicated.  For both loops, red line-segments 
indicate the minimum height that an AR loop system would need to achieve to be visible from Earth.
As the cartoon indicates, our model that assesses whether an AR is to be labeled ``Far-Side'' or 
``Occulted'' depends on the AR size and location.} 
\label{fig:vis}
\end{figure}

Related to the location of an AFT-AR is a ``visibility'' parameter        
assigned to the region (Figure~\ref{fig:vis}, Table~\ref{tbl:vis}).
On-disk ({\tt ``Disk''}) ARs are fully visible from Earth-side viewing and have
more than 95\% of the AFT-AR boundary falling within the AFT assimilation window that 
effectively cuts off at $\mu\approx 0.1$ ($\theta\approx84^\circ$).
Occulted or Limb ARs (``{\tt Occ}'') will either (1) have more
than 5\% of its AFT-AR boundary beyond the AFT assimilation window or (2) be located fully beyond
the visible limb but still at a location at which, given its size, the region's associated
coronal loops would be expected to be visible from the Earth during an
energetic event (Figure~\ref{fig:vis}).    The first boundary thus assigns {\tt Occ} to
regions that may be visually detectable from Earth but that are at extreme enough viewing
angles so as to generally be excluded from standard flare-forecasting methods
\cite<see discussions in>{allclear,ffc3_1,ffc3_3}.   Far-side or invisible ARs (``{\tt Far}'')
are fully invisible from the Earth.

\begin{table}
\caption{Visibility Labels}
\label{tbl:vis}
\begin{center}
\begin{tabular}{|c c p{8cm}|}
\hline
Moniker & Assignment & Description \\ \hline
``Disk'' & On-disk & Fully visible from Earth, $<5$\% of area beyond AFT assimilation window \\
``Occ'' & Occulted, Limb & $>5$\% area beyond AFT assimilation window, but still expected to be visible due 
to size and location (Eqn.~\ref{eqn:vis1}) \\
``Far'' & Far-side, Invisible & Not visible from Earth due to size and location\\
\hline
\end{tabular}
\end{center}
\end{table}

The latter boundary is not fixed; we establish whether
or not a simple semi-circle of a size based on the half-width (in longitude) of 
the AFT-AR and centered at the region's MCOM location, would
extend beyond the visible limb.   What is calculated first is essentially the minimum 
height that could be visible from the region's Stonyhurst longitude $\theta_{\rm Stony}$:
\begin{equation}
h_{\rm min} = R_\odot (\frac{1}{\cos{\theta_p}} - 1.0)
\label{eqn:vis1}
\end{equation}
where $\theta_{p} = \theta_{\rm Stony} - 90.$ is the distance (in longitude degrees) beyond the limb.
Then, if the semicircle that subtends the region's size has $h > h_{\rm min}$, the region is denoted as 
``Occulted'', otherwise it is a fully invisible ``Far'' AR.

\subsubsection{Magnetic Field Characterization}
Throughout, only pixels greater than an estimate of the AFT-map noise
are included in the analysis that produces the parameters.  
Based on the standard deviation in weak-signal areas this is around 15\,Mx\,cm$^{-2}$,
whereas the AFT-AR identification and tracking necessarily uses a higher
threshold.  Thus, only using pixels within the AFT-AR ``blobs'' (see Fig.\,\ref{fig:aft_eg}) 
for the parametrization module ensures that we only use pixels within the AFT-AR delineation 
and thus significantly above the noise level.

As this is a ``proof of concept'', we limit the number of magnetic field parameters 
that we present here to two, and use them together for multi-parameter analysis.  

The first parameter is the total magnetic flux of the region.  As discussed in 
\citeA{flareprediction,Welsch_etal_2009,allclear}, the total magnetic flux is an extensive 
parameter (related to the AR's size) that is well-known to be a fairly good predictor of 
flare activity and
to some extent can act as a benchmark against which to compare more sophisticated analysis. As such, 
the total unsigned magnetic flux is simply
$\Phi = \sum_{i=0}^N \left| B_{\rm r,i} \right| A_{\rm i}$ 
where $B_{\rm r,i}$ is the radial
magnetic flux density in each ${i}^{\rm th}$ pixel (Mx\,cm$^{\rm -2}$) over the 
$N$ pixels in the AFT-AR blob, and $A_{\rm i}$ is the pixel size in cm$^2$.

A second parameter that is available from just the $B_{\rm r}$ distribution but which also 
performs well on large-sample evaluations is the $\mathcal{R}$ parameter of
\citeA{Schrijver_2007}), or the total magnetic flux in the immediate vicinity of strong-gradient polarity-inversion lines (SGPIL).  
The polarity inversion lines are identified using a method similar to that described in
\cite{Schrijver_2007}, essentially masking each polarity individually, growing these masks, 
and identifying where they overlap as the SGPIL.  We use a 100\,Mx\,cm$^{-2}$ threshold and a 2-pixel 
($\approx 8.5$Mm) width; additionally we erode and grow the resulting map to remove 
single-pixel areas, and finally extend the boundaries by $\approx2$ pixels.  
This final area is the vicinity 
of ``strong-gradient PILs'' (s), noting that (1) there may be more
than a single SGPIL across a magnetic region, and that (2) we intentionally take
a small area around it, since a PIL is by definition a width-less line.  
The specific values and thesholds used here differ from the original \cite{Schrijver_2007},
but recover the intended morphology given the different data source, sensitivity, data grid, and 
spatial resolution.
We then calculate the area of the AFT-AR that satisfies these criteria,
and the flux contained within it.

It does happen that AFT-ARs will be centered close to the Carrington Longitude$=[0^\circ,360^\circ]$
boundary.  In this case, the analysis module identifies the situation, and creates
a new map with the longitude ``seam'' at Longitude=$180^\circ$, wrapping the AFT-map
and thus enabling the split region to be parameterized as a whole.

\subsection{Statistical Approach for $4\pi$ Flare Forecasting}
\label{sec:stats_general}

For this proof of concept study, we focus on statistical classification of 
labeled data.   As mentioned above, fully validating on far-side flare activity
is beyond the scope here; we focus on the ability to perform limb-flare
region classification with, {\it vs.} without, relevant region information.

\subsubsection{NWRA Classification Infrastructure}
\label{sec:nci}

The statistical evaluations performed here deploy the NWRA Classification Infrastructure
(NCI), which is the ``research arm'' of the operationally-running 
NWRA Discriminant Analysis Flare Forecasting System \cite<DAFFS; >[]{nci_daffs}.
NCI is a well-established statistical classifier facility that has served, 
both in its present form and during its earlier development, as the
statistical analysis tool for numerous investigations 
\cite{dfa,SWJ,dfa2,dfa3,Komm_etal_2011a,trt_emerge3,Welsch_etal_2009,Barnes_etal_2017,LinKusanoLeka_2021,nci_aia}.
Based on nonparametric discriminant analysis \cite<NPDA; >[]{ken83}, NCI uses non-parametric estimates 
of the probability density functions from samples of known populations to assign membership for a new
measurement, with cross-validation. In its basic form, DA produces a categorical prediction.  By using Bayes'
theorem, this is extended to produce the probability that a measurement belongs to one of two
(or more) populations: if ${\bf x}$ belongs to population $j$, then
\begin{equation} \label{eqn:prob} 
P_j({\bf x}) = {q_j f_j({\bf x}) \over q_1 f_1({\bf x}) + q_2 f_2({\bf x})}.
\end{equation}
\noindent
where $q_j$ is the prior probability of belonging to population $j$, $f_j({\bf x})$ 
is the probability density function for population $j$, and for this study $j=2$
refers to the event population, while $j=1$ refers to the non-event population.

From these density estimates and probabilities, NCI outputs a broad
range of evaluation data and metrics to quantify the ability of the chosen 
input parameter(s) to discriminate between the populations (see Section\,\ref{sec:validation}). 
For the nature of this study, we are not invoking bootstrap estimates of the uncertainties
at this juncture, but estimate skill-score uncertainties based on experience with similar 
sample sizes \cite{nci_daffs,nci_aia}.  

The needed components for NCI input are parameters to evaluate 
(Section\,\ref{sec:parameters}), and the labels for those parameters 
which takes the form of an ``event list'' (Section\,\ref{sec:eventlist}).
The NCI code package is designed to accommodate project-specific needs; in the
present case, the relevant modules were modified as per the specifics 
of the AFT-map parameters in particular the visibility labels.

\subsubsection{Event-List Generation}
\label{sec:eventlist}

The event lists construct the labels for the samples to be evaluated by means of the nonparametric 
density estimates. What is needed for this project is the time and location of a flare start, and an eventual 
peak magnitude of that flare.  The initial data source is the NOAA edited flare event
lists \cite{noaa_events}, which reports flare parameters based primarily on the Geostationary
Operational Environment System X-Ray Sensor \cite<GOES/XRS; >[]{goes_xrs}, in particular
the $1 - 8$\AA\ Soft X-Ray sensor.  For this study, we do not perform any further 
background subtraction or calibration.

The NOAA-provided flare lists do include some errors and missing information.  In 
particular, a number of flares are un-matched to NOAA~AR Numbers, or are missing
location information.  Missing information is especially common for flares that 
occur just behind the solar limb.   We performed a curation of the dataset with 
particular attention to these events.  A listed event for which a NOAA~AR was assigned
but a location was not, was confirmed and a location assigned using NOAA Region Lists 
and standard differential rotation rates, including propagating beyond-limb locations.  
Beyond-East-limb events are assigned to a ``future-assigned'' NOAA~AR according to date, 
location, and backwards-propagating differential rotation.  If there is no location or 
NOAA~AR number, we manually searched for a coinciding event using numerous 
available resources including NOAA/SWPC H$\alpha$ flare event listings, 
Solar Monitor \cite{solarmonitor}, the {\it SolarSoft} ``Latest Events Archive''
\cite{SSW,SSW_LatestEvents}, 
and other on-line public resources.  All assignments that involved such sleuthing 
are, of course, subject to interpretation; the curated list is available with the 
supporting data for this paper \cite{HD_4pi}.  Relevant for the analysis below, however,
is a recognition that Earth-viewpoint detection of {\tt Occ} flares will
be necessarily undercounted due to attenuation effects.

Thus, the AFT-ARs are linked to the flare events through their association with the NOAA AR 
numbering, as described above.   AFT-ARs may persist for multiple 
solar rotations, and therefore may be associated with more than one
NOAA AR.  Region and event matching was done via Carrington longitude and latitude, and by date.
At this juncture,
we are not attempting to incorporate truly far-side events as may be
observed by, {\it e.g.}, the Spectrometer and Telescope for Imaging X-rays
(STIX; \citeA{STIX,STIX_datacenter}) on board Solar Orbiter; such a 
large task is beyond the scope of this initial proof-of-concept study.

We generate lists based on two thresholds that defined an event, at the 
peak-flux levels of $10^{-6}\,{\rm W\,m}^{-2}$ (C1.0) and $10^{-5}\,{\rm W \,m}^{-2}$ (M1.0), 
with a validity period of 24\,hr starting at 00:00, 08:00 and 16:00UT.  
The list associates an event (or lack of event)
within the validity period for each AFT-AR.  ``Full-disk'' event lists are
then generated simply by tabulating whether any event occurred within any {\tt Disk} {\it or} 
{\tt Occ} AFT-AR.   Of note, the AFT maps are sampled three times per day with 8\,hr 
of evolution between each sample which we argue provides for derived parameters that
are independent;
the final relevant sample sizes are given in Table~\ref{tbl:flares}.

\begin{table}
\caption{Region and Flare Sample Sizes 2010.06 -- 2016.12}
\label{tbl:flares}
\begin{center}
\begin{tabular}{|l c c|}
\hline
\multicolumn{3}{|c|}{Region-by-Region} \\
 Sample Description  & On-Disk Regions & Occulted Regions \\
\hline
 AFT-ARs (``HMI Baseline'') & 23,485 & 4,825 \\
 \hspace{0.5cm} C1.0+ flares & 5,929 & 302 \\
 \hspace{0.5cm} M1.0+ flares & 1,012 & 67 \\
 AFT-ARs (``HMI + GONG '') & 23,892 & 6,580 \\
\hspace{0.5cm} C1.0+ flares & 5,923 & 522 \\
\hspace{0.5cm} M1.0+ flares & 1008 & 127 \\ \hline
 \multicolumn{3}{|c|}{Full-Disk} \\ 
 Sample Description & \# Samples & \#C1.0+ Flares/ \# M1.0+ Flares \\ \hline
 AFT-ARs (``HMI Baseline'') & 6,911 & 3,895 / 950  \\
 AFT-ARs (``HMI + GONG '')  & 6,944 & 3,995/ 1002 \\
\hline
\end{tabular}
\end{center}
\end{table}

\subsubsection{Validation Statistics}
\label{sec:validation}

The NPDA produces a probability that a data point (from the
parametrization) belonging to one of the assigned populations.
In this case, we invoke two populations - flare-imminent and
flare-quiet at the levels and in the validity periods described in
Section\,\ref{sec:eventlist}.  One evaluation metric reported here is
the Brier Skill Score (essentially a Mean Square Error Skill Score)
which summarizes Reliability Plots that graphically show a probabilistic
classification's reliability, skill, and resolution \cite<see>[for
discussion]{allclear,ffc3_1}.  The Reliability plots and BSS metric
do not rely on setting a threshold ($P_{\rm th}$) at which a
probability becomes a ``predict event'' {\it vs.}``predict no-event'',
in order to populate a ``truth table'' of True Positives, False negatives,
{\it etc.}

We also produce Receiver (Relative) Operating Characteristic
(Curve) (ROC) plots that compare the False Alarm Rate to the Probability of Detection,
essentially the components of the ``truth table'' as the $P_{\rm th}$ is 
stepped between [0.0, 1.0].  The ROC Skill Score (ROCSS, see \citeA{ffc3_1} for discussion)
summarizes the ROC plot with a normalized area under the curve such that ROCSS=0.0 corresponds 
to no skill, and ROCSS=1.0 is perfect skill.  As such, the BSS and ROCSS provide comprehensive 
information without being sensitive to an assigned $P_{\rm th}$.

The Peirce Skill Score (also known as the True Skill Statistic (TSS);  
\citeA{allclear,Bloomfield_etal_2012,ffc3_1}) is a dichotomous skill score 
popular in the literature.  It has been shown to have a maximum when the probability 
threshold is equal to the event rate, or  $P_{\rm th} = R = N_{\rm event}/ N_{\rm total}$ 
\cite{Bloomfield_etal_2012,Kubo2019}.  The location of max(TSS) on the ROC curves is 
indicated; it is the location of the maximum vertical distance from the $x=y$ ``no skill'' line
in the ROC plots.  The max(TSS) scores here are all ``reasonable'' (ranging 
roughly 0.45 - 0.6) but at some level, TSS and other dichotomous skill scores are distracting 
due to their sensitivity to $P_{\rm th}$.  

In this investigation, the climatological rate is an interesting challenge
due to the presumed attenuation of detected flares just beyond the solar
limb \cite{Woods_etal_2006}, and the inability to accurately account for
$N_{\rm total}$.  The event rate $R$ for {\tt Disk} is the most accurate
since all information should be present and accounted for.  Thus, all
climatological event rates applied during evaluation are $P_{\rm th} =
R_{\rm Disk}$.

Of interest here are the missed forecasts for events that could be Earth-impacting, 
{\it i.e.} the ``False Negatives'', thus we include in our validation statistics the change 
in FN entries (``$\Delta$FN'') between the test cases ({\tt F11} {\it vs.} {\tt F10}
or {\tt FH11} {\it vs.} {\tt FH10}) when 
$P_{\rm th} = R_{\rm Disk}$. For completeness but without dwelling on results that
are sensitivty to $P_{\rm th}$, we also report the change in ``False Positives'' (false alarms)
between the tests.  Finally, we also report the number of {\tt Occ} flares correctly 
predicted, or ``True Positives'', when limb information is included (the {\tt F11} 
and {\tt FH11} tests), for comparison to zero (necessarily) when no information is 
present (the  {\tt F10} and {\tt FH10} tests), again when $P_{\rm th} = R_{\rm Disk}$.  

We refrain from reporting numerous other skill scores in detail,
since for this study the emphasis is not on the performance of the
classification itself (as long as it is reasonable, and not pathologic),
but on evaluating improvements gained by having additional information
beyond Earth- facing sources.

\subsection{The Tests}
\label{sec:tests}

Tests are designed to evaluate the ``Delta'', or gain (if any) in the
statistical classification performance enabled by the availability
of near- and beyond-limb magnetic field data in the context of near-
and beyond-limb flares.  The different test configurations is given in
Table~\ref{tbl:tests}, with boolean indicators as to whether the occulted
regions are included in the density estimates (the ``training'').

For the tests, all event lists were the same in terms of the flare events
included, according to the threshold as indicated.  In other words,
if a flare occurred beyond the limb (but was detected in SXR by GOES),
it was included as an 'event'.  It is extremely likely that the smaller
flares from larger regions beyond the limb are under-sampled, presenting
a bias.  However, as the eventual goal of any flare forecasting approach
is to forecast for impactful events, we use the GOES-based detection of
the flares as a ground-truth in this study.

\begin{table}
\caption{Summary of Tests}
\label{tbl:tests}
\begin{center}
\begin{tabular}{lcccl}
\hline
Moniker & Disk  & Occulted  & Data / Region  & Probability \\
   &  Regions? & Regions? & Input & Basis \\
\hline
F10 & Y & N & Surface Flux Transport only & Region-by-Region \\
F11 & Y & Y & '' ''  & '' ''\\
F10-FD & Y & N & '' '' & Full-disk  \\
F11-FD & Y & Y & '' '' & '' '' \\
FH10  & Y & N & Surface Flux Transport + Helioseismology  & Region-by-Region \\
FH11  & Y & Y & '' '' & '' '' \\
FH10-FD  & Y & N & '' '' & Full-disk \\
FH11-FD  & Y & Y & '' '' & '' '' \\
\hline
\end{tabular}
\end{center}
\end{table}

\subsubsection{Disk Regions {\it vs.} Occulted Regions}
\label{sec:test_vis1}

The basic approach is to compare the NCI classification performance according to whether
or not the occulted regions are included when ``training'' the system -- {\it i.e.} computing the 
nonparametric density estimates upon which the region-based event probabilities are computed.  
The moniker indicates the boolean for [{\tt Disk},{\tt Occ}] regions included,
that is ``{\tt 10}'' {\it vs.} ``{\tt 11}'' in Table~\ref{tbl:tests} indicate not including
and including the {\tt Occ} regions, respectively.  

In other words, for all {\tt *10} tests, the region probabilities 
are computed using only the on-disk data and on-disk outcomes, duplicating the general approach 
of most forecasting facilities 
\cite<see discussions in >[]{allclear,ffc3_1,ffc3_2}.  In contrast, all {\tt *11}
 tests have region probabilities 
computed according to both the parameters (AFT-ARs) and outcomes, 
meaning on-disk ({\tt Disk}) and occulted
({\tt Occ}) parameters are used and both on-disk and occulted flares are used to 
compute the region probabilities.

For the evaluation, however, the event lists against which all {\tt *10}, and {\tt *11}
tests are evaluated
include all detected and region-assigned flares.  For the ``{\tt 10}'' tests, {\tt Occ} 
AFT-ARs are assigned a probability of 0.0 throughout.  In this test, then, any {\tt Occ} flare would 
be registered necessarily as a ``miss'' although {\tt Occ} non-event outcomes 
are registered as ``true negatives''.
This is not {\it exactly} the same as having ``no information'' that a region is present,
however many systems would assign a climatological prediction to a region about which 
there is otherwise no quantitative information, which would be generally $\gg\!0$ and thus
in fact directly indicates some knowledge of a region's existence.  In the scheme
adopted here, we are assured that occulted regions in the ``{\tt 10}'' tests will register
a ``miss'' or False Negative, but there is a bias towards True Negatives.

\subsubsection{Surface Flux Transport Only {\it vs.} Surface Flux Transport + Helioseismology}
\label{sec:test_helioseismology}

Tests are constructed to evaluate the additional performance enabled 
by incorporating far-side information from helioseismology (Table~\ref{tbl:tests}).
These maps from which the parameters are calculated appear otherwise indistinguishable, 
but underlying them are either the ``AFT-Baseline'' AFT maps which assimilate data only 
from HMI into the flux transport (``F'' monikers in Table~\ref{tbl:tests}), or the 
``AFT+GONG'' maps that include AFT-ARs detected and tracked from the GONG helioseismic data 
(``FH'' in Table~\ref{tbl:tests};  see Section\,\ref{sec:4piSun}).  
 
As is expected, there are just over 2,000 additional data points (AFT-ARs at any given target 
time every 8\,hr) over the 6.5-year period in the FH11 dataset over the F11, as the additional 
helioseismology information provides additional AR detections (Table~\ref{tbl:flares}).  The impact
of this additional information is clearly seen in the number of flare-positive AFT-AR data points, 
which are significantly more for the AFT+GONG than the AFT Baseline (for both C1.0+ and M1.0+)
for the occulted areas {\tt Occ}; the on-disk flare-positive AFT-AR data points
are essentially equal, also as expected given the data assimilation.  

\subsubsection{Region-by-Region {\it vs.} Full Disk}
\label{sec:test_fd}

The region-by-region tests compare directly the probability of an event 
being produced by a particular region with the outcome, as described above.  
The sample sizes are decently large and this is the manner by which many
forecasting algorithms work.  In practice, however, the location of the flare itself 
is irrelevant \cite<unless it initiates a forecast for related phenomena such as 
energetic particles;>[]{Whitman_etal_2023}.  
Additionally, full-disk forecasts can enable methodology
comparisons when the definitions of an AR differ \cite{ffc3_1}.

Hence, we conduct tests using a 
``full-disk approach'' \cite<>[Appendix
B]{ffc3_1}, meaning Sun-as-a-star-like, or agnostic of flare
location of the originating AFT-AR.  The full-disk probabilities are
constructed for each time from the AFT-AR probabilities \cite<see >[]{ffc3_1}:

\begin{equation}
\label{eqn:fdprob}
P_{\rm FD} = 1.0 - \prod(1.0 - P_{\rm AFT\_AR})
\end{equation}
where $P_{\rm AFT\_AR}$ is each AFT-AR's probability at a given time, and $P_{\rm FD}$ is the 
full-disk probability of an event.  As with the region-by-region tests, the {\tt F10-FD}
and {\tt FH10-FD} tests 
comprise probabilities for the on-disk AFT-ARs computed only against the on-disk events, and 
occulted AFT-ARs are assigned $P_{\rm AFT\_AR} = 0.0$ prior to computing the full-disk 
probabilities.  In this manner, the occulted AFT-ARs do not add to the full-disk event probabilities
for the {\tt F10-FD} and {\tt FH10-FD} tests.  On the other hand,
for both the {\tt F11-FD} and {\tt FH11-FD} tests, they do. For all {\tt -FD} tests, 
the evaluation is performed using the full event list but (where applicable), $P_{\rm th} = R_{\rm Disk}$.
As such, there
are 33 additional full-disk samples in the GONG-supplemented AFT-maps
that can be analyzed in the time-period because at least one ``{\tt Occ}'' region is
identified where none are present in the AFT Baseline maps.

While in practice some operational forecasting systems treat ``no/missing data''
or ``bad data'' by assigning a climatology-based forecast \cite{nci_daffs,ffc3_1} - 
in this case we strive to demonstrate the stringent scenario of the differences between 
having information {\it vs.} not.  Such a treatment also provides the most interpretable results.

\section{Results}
\label{sec:results}

The results presented here are in the spirit of a ``proof of concept''.  
Little effort has been made to achieve the best possible performance, 
beyond deploying two parameters that have been shown to have relevance 
for solar flare productivity.

\subsection{A Tale of Two Limb Flares}
\label{sec:cases}

So, did it work?  We look first at two examples, one a classification success and the other a classification failure.

\subsubsection{Success: NOAA~AR~12192 in October 2014}

NOAA AR\,12192 transited the disk mid-late October and displayed significant flare
activity.  The flares started before it was actually visible, however.  On 2014.10.14 at
08:00\,UT and then again at 16:00\,UT, the probability of an M1.0+
flare using the AFT+GONG ({\tt FH-11}) test 
was $>\!0.1$ (significantly larger than the $R=0.043$ climatology).  At this point the AR was 
centered at $-125^\circ$ 
longitude, or $35^\circ$ beyond the limb, but produced two GOES-detected M-class flares (M1.1 at 18:37\,UT 
and M2.2 at 19:07\,UT).  The probabilities stayed high, and indeed, AR\,12192 is a notable flare-active 
AR for Solar Cycle 24.

In contrast, the AFT-Baseline test that includes no far-side information had a probability 
of $\approx 0.01$ for both of those 2014.10.14 times quoted above.  The probabilities increased upon HMI data assimilation, 
but by that time the region had flared numerous times.

\subsubsection{Failure: NOAA~AR~12205 in November 2014}

This particular AR grew and diminished repeatedly 
during its lifetime with varying flare productivity, according 
to multiple analysis efforts, and proved to be a significant challenge. 
A new GONG-identified AFT-AR appears in the AFT+GONG maps at the end of October 2014
(AFT-ID {\tt 20141027TN21014}, {\it i.e.} emerging 2014.10.27 at N12 014 Carrington Longitude)
at almost $-180^\circ$ Stonyhurst longitude.  This AFT-AR then disappears on 2014.10.31 at 16:00\,UT  
near $\approx -130^\circ$ Stonyhurst longitude.  
As this Carrington longitude rotates into view and the HMI data-assimilation begins,
the magnetic flux concentrations grow starting 2014.11.05 and it acquires a new AFT-ID 
{\tt 20141108TN15010}  -- but then it again decreases in size a few days after that.  
As such, unfortunately, 
the large and loop-visible {\tt Occ} M-class flares on 2014.11.05 were classification misses.
Finally, after 2014.11.06 (when the AFT-AR and now NOAA~AR~12205 was at $\approx{\rm E}60$ Stonyhurst 
and fully a {\tt Disk}) AR, the subsequent multiple large flares were correctly classified.

Indeed an analysis using different approach \cite{Hamada_etal_2024,Hamada_etal_2025} 
corroborated that the seismic signature in this location diminished between 
2014.10.27--2014.10.29. This particular AR may prove to be a good test-case for 
future improvements to helioseismic detection of near-limb ARs.

\subsection{Large-sample Analysis}
\label{sec:largesample}

\begin{figure}[t]
\centerline{
\includegraphics[width=1.0\textwidth,clip, trim = 0mm 0mm 0mm 0mm, angle=0]{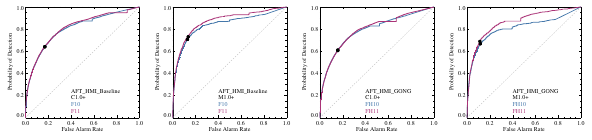}}
\caption{ROC plots for the region-based forecasts, comparing the paired tests as indicated 
({\it c.f.} Table~\ref{tbl:tests}). 
For each, the $P_{\rm th}$ location on the ROC curve is indicated, where the TSS = POD-FAR
is maximum, is indicated.  The associated ROCSS and the differences between the paired 
tests are provided in Table~\ref{tbl:results_region}.}
\label{fig:roc}
\end{figure}

\begin{figure}
\centerline{
\includegraphics[width=1.0\textwidth,clip, trim = 0mm 0mm 0mm 0mm, angle=0]{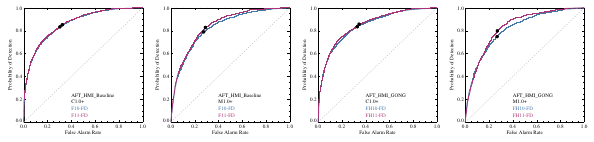}}
\caption{Same as Figure~\ref{fig:roc} but for full-disk forecasts, with associated ROCSS
given in Table~\ref{tbl:results_FD}.}
\label{fig:roc_FD}
\end{figure}

\begin{figure}
\centerline{
\includegraphics[width=1.0\textwidth,clip, trim = 0mm 0mm 0mm 0mm, angle=0]{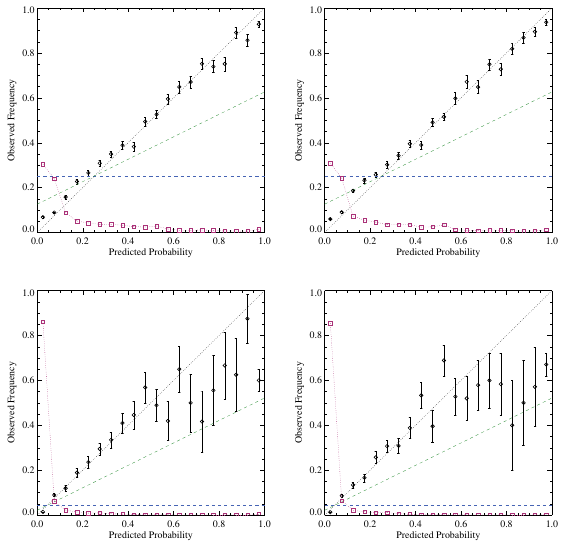}}
\caption{Reliability Plots for the AFT-HMI Baseline data without (F10, left) and with (F11, right), 
the occulted-region data included, for C1.0+ flares (top) and M1.0+ flare (bottom).  
The predicted probabilities are divided into 20 bins, and the error bars reflect 
the sample size of that bin \cite{Wheatland2005}.  Also shown are the sample-size
histograms (red), the climatology (blue) and the ``no-skill'' line (green).
}
\label{fig:reliability_base}
\end{figure}

\begin{figure}
\centerline{
\includegraphics[width=1.0\textwidth,clip, trim = 0mm 0mm 0mm 0mm, angle=0]{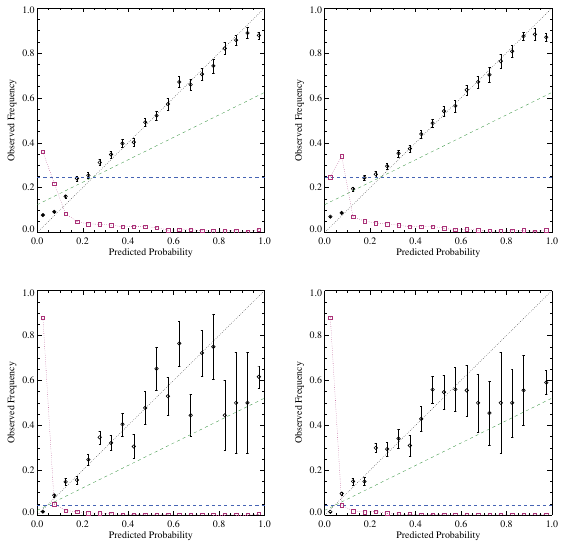}}
\caption{Same as Figure~\ref{fig:reliability_base} but for the AFT-maps
informed by the GONG-based helioseismology, {\it i.e.} FH10 (left) and FH11 (right).}
\label{fig:reliability_gong}
\end{figure}

\begin{figure}
\centerline{
\includegraphics[width=1.0\textwidth,clip, trim = 0mm 0mm 0mm 0mm, angle=0]{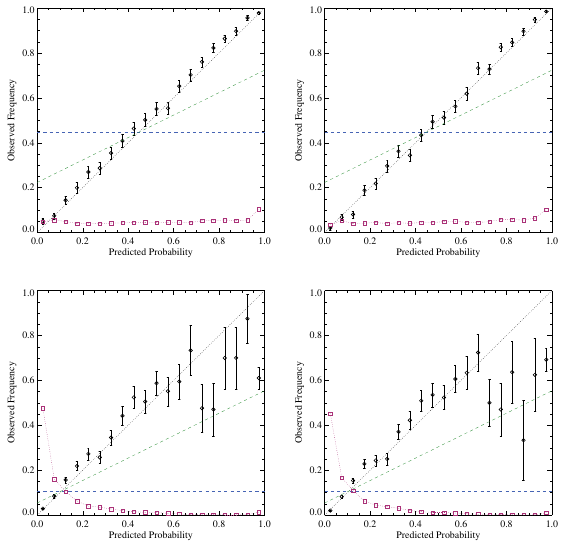}}
\caption{Same as Figure~\ref{fig:reliability_base} but for Full-Disk forecasts,
{\it i.e.} F10-FD (left) and F11-FD (right).}
\label{fig:reliability_base_FD}
\end{figure}

\begin{figure}
\centerline{
\includegraphics[width=1.0\textwidth,clip, trim = 0mm 0mm 0mm 0mm, angle=0]{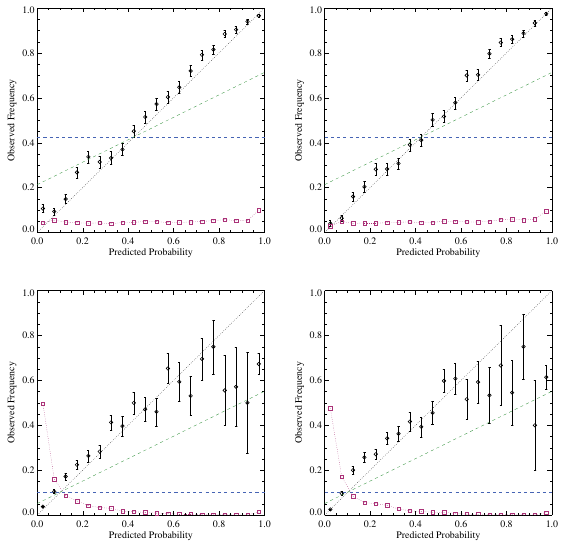}}
\caption{Same as Figure~\ref{fig:reliability_gong} but for Full-Disk forecasts,
{\it i.e.} FH10-FD (left) and FH11-FD (right).}
\label{fig:reliability_gong_FD}
\end{figure}

A summary of the full-sample tests and their resulting metrics are shown in Table~
\ref{tbl:results_region} for region-by-region tests and in Table~\ref{tbl:results_FD} for 
Full-Disk tests.  ROC plots are shown in Figures~\ref{fig:roc} -- \ref{fig:roc_FD}, and 
Reliability plots are shown in Figures~\ref{fig:reliability_base} -- \ref{fig:reliability_gong_FD}.

\begin{table}
\caption{Results Summary: Region-by-Region}
\label{tbl:results_region}
\begin{center}
\addtolength{\tabcolsep}{-0.25em}
\begin{tabular}{lccccccccc}
\hline
 Test & Event  & Climato-&$\Delta$(FN) & $\Delta$(FP) & TP ({\tt Occ}) & ROCSS & $\Delta$ROCSS & BSS & $\Delta$BSS  \\
  & Definition & logy ($R$)  & $P_{\rm th}\!=\!R$ & $P_{\rm th}\!=\!R$ & $P_{\rm th}\!=\!R$  &  &  &  \\
\hline
 F10 & C1.0+ & 0.252 & & & 0 & 0.606 & & 0.255 & \\
 F11 & '' '' & '' '' & -10 & -1 & 116  & 0.617 & +0.011  & 0.264 & +0.009 \\
 F10 & M1.0+ & 0.043 & & & 0 & 0.676 & & 0.161 &  \\
 F11 & '' '' & '' '' & -26 & +190 &  34 & 0.734 & +0.058 &  0.178 & +0.017 \\ \hline
 FH10 & C1.0+ & 0.248 & & & 0 & 0.574 & & 0.232 & \\
 FH11 & '' '' & '' '' & +31 & -9 &  155 & 0.584 & +0.010 &  0.240 & +0.008 \\
 FH10 & M1.0+ & 0.042 & & & 0 & 0.615 & & 0.149 & \\
 FH11 & '' '' & '' '' & -24 & -111 & 60 & 0.718 & +0.103 &  0.151 & +0.002 \\
\hline
\end{tabular}
\end{center}
\end{table}

\begin{table}
\caption{Results Summary: Full Disk}
\label{tbl:results_FD}
\begin{center}
\addtolength{\tabcolsep}{-0.25em}
\begin{tabular}{lccccccccc}
\hline
 Test & Event  & Climato-&$\Delta$(FN) & $\Delta$(FP) & TP ({\tt Occ}) & ROCSS & $\Delta$ROCSS & BSS & $\Delta$BSS  \\
  & Definition & logy ($R$)  & $P_{\rm th}\!=\!R$ & $P_{\rm th}\!=\!R$ & $P_{\rm th}\!=\!R$ &  &  &  \\
\hline
 F10-FD & C1.0+ & 0.446 & & & 0 & 0.719 & & 0.383 & \\
 F11-FD & '' '' & '' '' & -73 & +62 &  15 & 0.723 & +0.004 & 0.394 & +0.011 \\
 F10-FD & M1.0+ & 0.108 & & & 0 & 0.656 & & 0.188 & \\
 F11-FD & '' '' & '' '' & -37 & +85 & 0 & 0.686 & +0.030 & 0.209 & +0.021 \\ \hline
 FH10-FD & C1.0+ & 0.426 & & & 0 & 0.687 & &  0.343 & \\
 FH11-FD & '' '' & '' '' & -90 & +45 & 49 & 0.702 & +0.015 &  0.364 & +0.021 \\
 FH10-FD & M1.0+ & 0.103 & & &  0 & 0.627 & & 0.174 & \\
 FH11-FD & '' '' & '' '' & -50 & +15 & 4 & 0.662 & +0.035 & 0.176 & +0.002  \\
\hline
\end{tabular}
\end{center}
\end{table}

Overall the ROCSS and BSS are respectable on their own, and as has been stated before, 
are not the focus here.  The BSS decrease with higher
flare threshold (M+) whereas ROCSS increase, as has been seen previously \cite{allclear,ffc3_1}.  
Both the ROC plots and Reliability plots on ``on par'' with those shown 
in numerous publications, confirming that the ``$4\pi$-framework'' is generally 
performing as expected.

The statistical improvements upon adding ``{\tt Occ}'' data are overall in the `direction' 
that is expected:  ROCSS and BSS
scores improve, although the magnitude by which they increase with additional information 
is not large, especially for the {\tt F-FD} and {\tt FH-FD} tests.  
We expect the region-based 
uncertainties in these skill 
scores to be $\simeq\!0.01$ (C+) and $\simeq\!0.02$ (M+) given the sample sizes
\cite{allclear,nci_daffs,nci_aia}.    As such, the improvements as seen by 
$\Delta$ROCSS are of order 1$\sigma$ for C1.0+ and over $3\sigma$
for M1.0+, but all in the $\approx 0.0\sigma - 1.0\sigma$ range for $\Delta$BSS.  

These minor but positive improvement trends are reflected in the plots as well.
The ROC plots for C1.0+ for {\tt F10} {\it vs.}, {\tt F11} and {\tt FH10} {\it vs.} {\tt FH11} 
tests show little difference, while
there is a distinct difference for M1.0+ ROC plots (Figure~\ref{fig:roc}) as less
scatter between probability bins.  The improvement
is more noticeable for the {\tt FH} test than the {\tt F}, which is reflected in the higher $\Delta$ROCSS
in Table~\ref{tbl:results_region}.  Any differences in the Reliability plots are difficult 
to gauge by eye -- as reflected in the $\Delta$BSS.  For {\tt FH-11} / M1.0+, one can see 
a slight improvement as a ``tightening'' around the $x=y$ line for lower-probability
bins, but a distinct over-forecasting especially at higher probabilities that is somewhat 
more pronounced than for the {\tt FH-10} test.  This trend is directly related to the 
under-representation of {\tt Occ} flares, meaning it is more an under-detection
than an over-forecast.

The full-disk tests ({\tt F-FD} and {\tt FH-FD}) show similar statistical results: 
all $\Delta$[ROCSS, BSS] 
are in the correct ``direction'', but the statistical improvements are small.  This result
is not unexpected given the overall small fraction of ARs and flares that reside in the ``{\tt Occ}''
regime. 

The ROC plots for the FD-tests are again minimally improved for C1.0+, 
noticeably improved (albeit a small improvement) for M1.0+, with the FH test
showing a larger visual improvement.  It is difficult to see trends in the 
Reliability plots, as expected from the $\Delta$BSS results, except for an 
improvement for the lower-probability bins in the FH11-FD compared with the FH10-FD 
for C1.0+.

Both region-based and full-disk-based tests generally show reductions in False Negatives, 
more significantly for the {\tt F-FD} and {\tt FH-FD} tests and except for 
the FH (region) case.  
The trend is encouraging, although the specific numbers are sensitive to $P_{\rm th}$
and the difference between $P_{\rm th}$ for {\tt Disk} and {\tt Disk+Occ} can be 
up to $25\%$ from the available event lists again due to the under-detection of events
for {\tt Occ} plus the unknown ``non-event'' rate. 
There are generally trade-offs within a truth-table between False Negatives and False Positives 
and this is seen in Tables~\ref{tbl:results_region}, \ref{tbl:results_FD}.
Depending on user preferences, $P_{\rm th}$ can be optimized for improved performance for one 
or the other error and at the $P_{\rm th}=R_{\rm Disk}$ indeed the False Positives 
may or may not decrease with the added {\tt Occ}-data information.

With those $P_{\rm th}$ caveats in mind, most telling are the ``TP ({\tt Occ})'' entries.
For all ``{\tt -10}'' tests these are necessarily ``0'', meaning that no {\tt Occ} event 
is successfully classified. For all but one of the ``{\tt -11}'' tests, ${\rm TP ({\tt Occ})} \gg 0$
showing that with the $4\pi$ information, near-, at-, and just-beyond-limb events were indeed
successfully classified.  For the {\tt FD} tests, the numbers are smaller than region-tests, but
one must remember that these represent essentially $4\pi$ evaluations
from the Earth vantage point: each TP ({\tt Occ}) entry represents
a 24\,hr period (recalling our 8\,hr sampling) when the {\it only} event was a {\tt Occ} event, 
and that without the $4\pi$ information provided, that event would have been missed.
Moreover, the AFT-GONG {\tt FH} tests provide systematically more such cases than 
AFT-Baseline for both region- and FD-based tests, demonstrating that the
additional information provided by the far-side helioseismology can lead to better 
limb-ward flare classification.

The difference between East- and West-limb {\tt Occ} classification performance 
can be used to test the hypothesis that better outcomes are expected for the West-limb flares due to 
data being more recently-assimilated into the AFT model.  We present in Table~\ref{tbl:eastwest}
the results of the correctly-classified {\tt Occ} AFT-ARs separated by East-limb {\it vs.} West-limb.
The number of correct classifications uses the same $P_{\rm th}\!=\!R_{\rm Disk}$ as 
in Table~\ref{tbl:results_region}, and the ``Imbalance Ratio'' follows the Flux Balance 
(Eqn.~\ref{eqn:fluxbalance}), $H = \left | (N_{\rm East} - N_{\rm West})\right| / N_{\rm total}$ 
so that $H=0$ is perfectly
balanced and $H=1$ is perfectly imbalanced with $N_{\rm East}, N_{\rm West}$ being the 
true positives for East- and West-limbs respectively.
This hemisphere-focused analysis confirms the stated hypothesis when solely the AFT model
is used (the F11 tests), by way of high imbalance ratio and a dominance by $N_{\rm West}$.  
Importantly, the imbalance ratio significantly decreases when 
the far-side seismology information is included (the FH11 tests).  It is clearly demonstrated
thus, that the $4\pi$ framework improves limb-flare classification, and that there is significant
value-added particularly for East-limb flares by including information provided by 
far-side helioseismology.

\begin{table}
\caption{East {\it vs} West ``{\tt Occ}'' Success (Region-by-Region)}
\label{tbl:eastwest}
\begin{center}
\begin{tabular}{lccccc}
\hline
 Test & Event Def.& Climatology & $N_{\rm East}$ & $N_{\rm West}$ & Imbalance (H) \\
F11 & C1.0+ & 0.252 & 8 & 108 & 0.86\\ 
FH11 & C1.0+ & 0.248 &  45 & 110 & 0.42 \\
F11 & M1.0+ & 0.043 & 2 & 32 & 0.88 \\
FH11 & M1.0+ & 0.042 & 24 & 36 & 0.20 \\
\hline
\end{tabular}
\end{center}
\end{table}

\section{Discussion and Conclusions}
\label{sec:disc}

We present here a proof-of-concept ``$4\pi$ Solar Energetic Event
Forecasting System'', beginning with solar flares.  In particular,
we present this infrastructure looking forward to the need for
full-heliosphere space-weather forecasts as more missions populate the
solar system beyond the immediate Sun-Earth line.  For the near-term we
focus on the prospect of improving forecasts for Earth-impacting solar
flares that occur near or just beyond the solar limb (the ``Limb-Flare
Challenge'') where the data usually relied upon to make these forecasts
becomes unreliable or simply unavailable.

The new $4\pi$ framework consists of a surface flux transport model,
information input from far-side helioseismology, and a statistical
classifier.  Additional components developed for this system include
an appropriate flare event list (in this case, curated especially to
characterize near- and beyond-limb flares), a new AR-identification and
labeling scheme that mitigates problems brought by Earth-side sequential
numbering schemes, and a validation methodology that evaluates performance
statistically in the context of known detection bias present for this
challenge.

Tests were performed for a statistically-significant sample size covering
6.5 years (numerous solar rotations), for both region-by-region and
full-disk representations, following standard flare-forecasting event
definitions.  
Even with this multi-year dataset, however, the sample sizes were not
large, which meant that the differences between including the limb
information and not could be statistically subtle.  We mitigated this
challenge somewhat by using overlapping validity periods (deploying 24\,hr
validity periods every 8\,hr) to increase the sample size.  Our argument,
that the evolution of the AFT maps between 8\,hr samples provides
independent observation-classification outcome pairs, may not completely
guard against bias due to non-independent event labels.  We have tested
the analysis using the data for the three times chosen separately ({\it
e.g.}, three different sets of single-time sampling and 24\,hr validity
periods thus providing three sets of fully independent 24\,hr labels
and outcomes), each of which having roughly 1/3 the sample size of the
``combined-data''.  We have confirmed that for the present experiments,
the 8\,hr sub-sampling results in no significant bias being introduced
to the probability distributions and that there are no statistically
significant differences in the event rates or resulting skill scores,
especially given the increased uncertainties expected from the smaller
sample sizes.  The primary differences seen between single-time results
and the combined-data results are slightly improved skill scores for
the latter due to the larger sample sizes and thus better determination
of the probability density functions.  The data by which to confirm
our conclusions are available in the associated open-data repositories
\cite{HD_4pi}.

As such, our experiments find that information provided from solely the
surface magnetic flux transport model can, indeed, improve the forecasts
of Earth-visible limb flares according to standard validation metrics.
In particular, this framework can reduce the number of ``missed'' events
overall, and provide successful classification of West-limb-flares
specifically.  We find further improvement, in particular for
East-limb flares, when far-side detection of new solar ARs or the growth
of previously-identified regions are incorporated into the model using
far-side helioseismology.

Arguably, the improvements we show are statistically small when evaluated over 
all visible flares.  However, they
are consistently in the expected direction, and demonstrate well this proof of
concept.  Every single piece of the process described here can be 
improved.  Similarly, for each component of the framework a different methodology 
or data source could be invoked.  We focus here on using data that are ostensibly 
available in an operational manner, meaning we do not need to rely on data from
{\it off} the Sun-Earth line.  

However, the new infrastructure presented here is notable, including: 
\begin{enumerate}
\item a novel AR numbering scheme that avoids Earth-centric confusion brought 
on by solar rotation, allows for post-facto changes and provides unique identifiers;
\item methodology to incorporate the detection of far-side ARs and an evaluation
    of their magnetic properties, into a surface magnetic flux-transport model of the $4\pi$ Sun; \\
\item characterizing the $4\pi$ Sun using magnetic-field parameters known to be relevant to 
solar flare production;
\item developing the validation methodology for near-limb and occulted regions that are
otherwise missed by the vast majority of flare-forecasting systems.
\end{enumerate}

When or if it becomes timely to pursue an operational
version of a $4\pi$ energetic-event forecasting tool, identifying the best-performing
magnetic flux-transport system, most sensitive seismic detection algorithm, the most
stable AR
identification and tracking system, and the optimal set of magnetic 
parameters should indeed all be considered.  Indeed, to achieve the best ``limb-flare''
predictions, one could consider combining this $4\pi$ magnetic field based approach
with one that focuses on UV coronal emission visible above the solar limb
\cite{nci_aia,LeeJ_etal_2024}.  However, for this proof of 
concept, the point is not to obtain the highest possible
skill scores.  Instead we present a novel approach, describe the needed
infrastructure that was developed, and demonstrate the utility of $4\pi$ ``Full
Heliosphere'' space-weather forecasting by initially addressing the ``Limb-Flare Challenge''
failure-mode in today's operational solar flare prediction.

\section*{Open Research Section}

This project comprises data and code from many different sources, some ``pipeline'' and/or public, and some created specifically for this work.  Listed roughly by the institution that provided the data and/or is the responsible party, the availability is as follows.

\noindent{\bf NSO}:  
\begin{enumerate}
\item Phase-shift maps are available at: \url{https://gong2.nso.edu/archive/patch.pl?menutype=farside#step2}
\end{enumerate}

\noindent{\bf SwRI}:  
\begin{enumerate}
    \item Baseline AFT maps are available, see \citeA{aftbaseline}.  The most updated AFT Baseline map can be also obtained from \url{https://data.boulder.swri.edu/lisa/AFT_Baseline/}, which is being updated daily with latest availble HMI magnetogram data.
    \item AFT maps with GONG-informed input are available, see \citeA{aftgongfs}.
    \item The AFT-AR code with which the AFT-ARs are identified, tracked, and labeled is available, see \citeA{autotab2025} 
\end{enumerate}

\noindent{\bf NWRA}:
\begin{enumerate}
    \item A project-specific repository has been created, see \citeA{HD_4pi}.
    In this
    repository we provide: IDL ``save'' files that include the AFT-AR parametrization data,
    and the event populations. These files contain summaries, essentially, of hundreds 
    of output files otherwise produced that are large and contain significant amounts of 
    data extraneous to this project.  The forecast, or classifier probabilities are provided for 
    each of the tests, and JSON files that summarize the AFT-ARs (provided by SwRI) are 
    also provided here. Also provided are the ``curated'' event lists (the information from 
    which informs the "POP" label in the IDL save files'' structures).
    \item NCI, by which probabilities are computed, is described in \citeA{nci_daffs}.
    The linear version of discriminant analysis is available at {\url{https://www.cora.nwra.com/~graham/DA.html}}.  
    Some project-specific compute-environment-specific data handling wrappers were
    developed and not included, as not being general analysis tools.
    \item The codes for computing skill scores, ROCSS, and that are needed to produce the plots are provided in the \cite{HD_4pi} repository. 
\end{enumerate}

\newpage
\acknowledgments

This project was funded primarily by NASA/R2O2R grant \#80NSSC22K0273 with additional 
support from NASA LWS/SC grant 80NSSC22K0892 (subaward to NWRA); 
KJ was partially supported by the NSF Windows on the Universe - Multi-Messenger Astrophysics 
(WoU-MMA) grant  to the National Solar Observatory.
All opinions expressed herein are solely those of the authors.
This work utilizes GONG data obtained by the NSO Integrated Synoptic Program, 
managed by the National Solar Observatory, which is operated by the Association of Universities 
for Research in Astronomy (AURA), Inc. under a cooperative agreement with the National Science 
Foundation and with a contribution from the National Oceanic and Atmospheric Administration. 
The GONG network of instruments is hosted by the Big Bear Solar Observatory, High Altitude Observatory, 
Learmonth Solar Observatory, Udaipur Solar Observatory, Instituto de Astrof\'{\i}sica de Canarias, and 
Cerro Tololo Inter-American Observatory. We acknowledge the HMI and SDO facilities and teams, 
as well as the GONG facility and team, especially Mitch Creelman, for making this project possible, 
Dr. Amr Hamada for input on NOAA AR~12205,
and Dr. Graham Barnes for support on NCI.  The authors also thank the two thoughtful referees who helped us clarify and strengthen the presentation.

\section*{Conflict of Interest Statement}  
The authors affirm there are no conflicts of interest.

\newpage




\end{document}